
\input jnl.tex
\input eqnorder.tex
\input reforder.tex
\def\submit                     
        {\vskip 24pt \beginlinemode
        \noindent \rm Submitted to: \sl}
\def\sk{\smallskip}
\def\ni{\noindent}
\def\sq{\it {sq}}
\def\sc{\it {sc}}
\def\bcc{\it {bcc}}
\def\fcc{\it {fcc}}
\def\s{\rm {s}}
\title Multi-interaction mean-field renormalization group
\bigskip
\author C. N. Likos$^{(a)}$ and A. Maritan$^{(b)}$
\affil
$^{(a)}$ Dipartimento di Fisica Teorica, Universit\`a di Trieste,
Strada Costiera 11, I-34014 Grignano (TS), Italy
\smallskip
$^{(b)}$ Istituto Nazionale di Fisica della Materia (INFM),
Scuola Internazionale Superiore di Studi Avanzati
and INFN, Sezione di Trieste, via Beirut 2-4, I-34013 Trieste, Italy

\abstract We present an extension of the previously proposed
mean-field renormalization method to model Hamiltonians
which are characterized by more than just one type of interaction.
The method rests on scaling assumptions about the magnetization
of different sublattices of the given lattice and it generates
as many flow equations as coupling constants without arbitrary
truncations on the renormalized Hamiltonian. We obtain good results
for the test case of Ising systems with an additional
second-neighbor coupling in two and three dimensions.
An application of the method is also done to a
morphological model of interacting
surfaces introduced recenlty by Likos, Mecke and Wagner
[J. Chem. Phys. {\bf{102}}, 9350 (1995)].
\bigskip
\ni
\submit {Physical Review E}
\endtopmatter

\head{\bf{I. Introduction}}
\taghead{1.}

One of the simplest techniques for deriving renormalization
group recursions is the so-called mean-field renormalization group (MFRG)
of Indekeu {\it{et al.}}\refto{mfrg} The MFRG has proven to be an efficient
method for the calculation of critical couplings and
exponents of a wide variety of lattice models. The range of applications
of the MFRG extends beyond the study of bulk critical phenomena in classical
systems, to include quantum models and surface criticalities; moroever,
extensions of the MFRG ideas to dynamical critical phenomena have also
been presented.\refto{{review}, {stella}}

The most appealing features of the MFRG are its simplicity and its
direct connection with the classical mean-field ideas; whereas mean-field
theory is a simple approach to the study of the phase behavior of various
models in statistical mechanics, its most serious drawback is that it
predicts the wrong (classical) critical exponents. The MFRG offers a way
to remedy this deficiency while maintaining some basic notions from the
mean-field approximation (e.g. the effective field.)

We shall now summarize the main ideas of the original MFRG.
To illustrate the method, let us consider the
nearest-neighbor Ising
model which is described by the Hamiltonian:
$$ H \equiv -\beta {\cal{H}} = h \sum_{i} s_i + J \sum_{<ij>} s_i s_j,
   \eqno(ising1) $$
where the second sum is carried over nearest-neighbor
sites on a given lattice of dimension $d$.
Consider now two clusters of interacting spins,
containing $N$ and $N'$ sites, with $N' < N$. Suppose that
the surrounding spins of these clusters is fixed at the values
$b$ $(|b| \leq 1)$ for the large cluster
and $b'$ $(|b'| \leq 1)$ for the small one and evaluate
the magnetizations per site for
both clusters, $m$ and $m'$ respectively. Clearly, $m$ and $m'$ depend
on the coupling constant $J$,
the magnetic field $h$ and the boundary values $b$ and $b'$.
Setting $m(J,h,b)=b$ or $m'(J,h,b')=b'$ we would obtain two different types of
mean-field equations, which could then be solved self-consistently and the
criticality would be identified with the `bifurcation point' of any of the
two equations for $h=0$ (each of the two equations yields a different
approximate value $J_c$ for the critical coupling, of course.)
However, this approach leads to classical critical behavior.
Instead, in the MFRG one uses the two cluster magnetizations to define a
{\it{mapping}} $(J,h,b) \to (J',h',b')$ by requiring
$$ m'(J',h',b') = {\s}^{d-y_h} m(J,h,b), \eqno(map1) $$
together with
$$ b' = {\s}^{d-y_h} b \eqno(scaleb1) $$
to hold to leading orders in $h$ and $b$. In Eqs. \(map1) and \(scaleb1)
above, the rescaling factor ${\s}$ is defined as
$$ {\s} = (N/N')^{1/d}. \eqno(defl) $$
Setting $h=h'=0$, expanding
both sides of
\(map1) to linear order in $b$ and $b'$ and using \(scaleb1) we obtain
the RG flows for the coupling constant in the form
$$ {{\partial m'(J',h',b')}\over{\partial b'}}\bigg|_{h'=0,b'=0}=
   {{\partial m(J,h,b)}\over{\partial b}}\bigg|_{h=0,b=0}. \eqno(flows1) $$
Once the fixed points $J_*$ of the iteration have been found
from \(flows1) the thermal and magnetic exponents $y_t$ and $y_h$ are
determined through the relations
$$ y_t = {{\ln \lambda_t}\over{\ln{\s}}}, \eqno(yt1) $$
where
$$ \lambda_t = \Bigl[
   \Bigl({{\partial^2 m(J,h,b)}\over{\partial b \partial J}}\Bigr)
   \Bigl({{\partial^2 m'(J',h',b')}\over{\partial b' \partial J'}}\Bigr)^{-1}
   \Bigr]\bigg|_{J=J'=J_*,h=h'=0,b=b'=0} \eqno(lt1) $$
and
$$ {{\partial m'(J',h',b')}\over{\partial h'}}\bigg|_{J'=J_*,h'=0,b'=0}=
   {\s}^{d-2y_h}
   {{\partial m(J,h,b)}\over{\partial h}}\bigg|_{J=J_*,h=0,b=0}. \eqno(yh1) $$
As a rule, the MFRG yields better estimates for the critical
coupling $J_c \equiv J_*$ than it does for the critical exponents.
However, the latter can also be significantly improved in a number of
ways. Slotte\refto{slotte} has shown that a redefinition
of the rescaling factor ${\s}$ leads to better values for the thermal
exponent than the `naive' definition of ${\s}$, eq. \(defl) above.
On the other hand, Indekeu
{\it{et al.}}\refto{indekeu} have succeeded in improving the MFRG critical
exponents in a more systematic way, by considering three clusters and
introducing an additional surface critical exponent $y_{h_s}$; this
approach has led to a unified approach to bulk, surface and corner
critical behavior (see also \ref{review}.)

The remaining of the paper is organized as follows: in Section II we
motivate and
present a new, extended version
of the MFRG. In Section III we test the
new approach by applying it to the square, simple cubic and body-centered
cubic Ising models with crossing bonds, obtaining very good results
for the fixed points and the overall phase behavior of these models.
In Section IV we then apply this RG scheme to
a morphological Hamiltonian of interacting surfaces.\refto{wagner}
In Section V we summarize and conclude.
Since the derivation of the RG
flows for the morphological model is
a bit lengthy, we outline the main steps
in the Appendix.

\head{\bf{II. Extended MFRG}}
\taghead{2.}

Let us consider, to begin with, what happens if we want to renormalize,
by means of the simple MFRG-scheme,
a Hamiltonian in which the even interaction part ({\it{i.e.}} excluding the
magnetic field and having a remainder which is invariant under
$s_i \to -s_i$)
contains more than just one type of interaction and thus
more than one coupling constant. Denoting by
${\bf{J}} \equiv (J_1,J_2,\ldots,J_\nu)$ the $\nu$-dimensional coupling
constant of the Hamiltonian, the MFRG which is based on the scaling of
a {\it{single}} order parameter (the bulk magnetization) always leads to
a {\it{single}} flow equation, namely one of the form \(flows1) with
$J$ and $J'$ replaced by their multidimensional counterparts
${\bf{J}}$ and ${\bf{J'}}$ respectively.
Clearly, in order to specify the flows one needs
as many equations as coupling constants in the
Hamiltonian, so this single equation is not sufficient.
In the vast majority of MFRG
applications to many-couplings Hamiltonians, this equation was used
in the following sense: a `fixed-point requirement' of the form
${\bf{J}}={\bf{J'}}$ was made, which led to an equation of the form
$f({\bf{J}})=0$, identified as the equation which defines the critical
surface of the model. Although such an approach has been widely
used,\refto{review} and gives qualitatively correct results, it has the
obvious drawback that it yields infinitely many `fixed points' (each point
on the critical surface is `fixed') and thus detailed information on
the critical exponents is lost in this scheme. Moreover, the approach only
serves to define the critical surface {\it{alone}} and not to determine
the way in which each of the individual couplings $J_{\alpha},\;
\alpha = 1,2,\ldots,\nu$ flows in the presence of the others.
Therefore, such an approach is useful only in the sense that it provides
us with better estimates of the critical surface of a given model
than the simple mean-field approximation, requiring roughly the same
amount of computational effort; but it does not offer a means for
the renormalization of the Hamiltonian.

There have been a few attempts to go beyond the above limitations of
the MFRG: de Oliveira
and S\'a Baretto\refto{oliv} studied the two-parameter
Hamiltonian of the Ashkin-Teller (AT) model\refto{at} and identified {\it{two}}
order parameters in the problem. Making scaling assumptions about
the order parameters of small and large clusters, they were then able
to obtain flow diagrams of the usual type, finding isolated fixed points,
critical exponents etc. However, the AT model is one in which {\it{two}}
different types of spin variables live on the lattice sites, and this
makes the introduction of two types of `magnetization' natural in this
model. An approach which has some similarities to ours was presented
by Plascak\refto{plas} for the case of the two-dimensional Ising model
with crossing bonds. This approach is based on the selection of
appropriate pairs of order parameters, depending on the region of the
parameter space considered. However, this method leads to some
ambiguities in the flows along the axis of vanishing first-neighbor
coupling.\refto{plas} On the other hand, as we show below, our approach
is free of such ambiguities.

We first present the basic conditions to be satisfied for the implementation
of the new method, at present. The types of models on which our approach
is applicable must satisfy the following requirements (the discussion of
more general cases is postponed for the concluding Section):
\sk
\ni
(i) the interaction part of the Hamiltonian must be even, and it must
contain only two coupling constants, call them $J$ and $K$;
\sk
\ni
(ii) the model must be defined on a bipartite lattice, and
\sk
\ni
(iii) there must be a region of the parameter space for which there are
only two kinds of ground states, a ferromagnetic (FM) and antiferromagnetic
(AFM) one, separated by some borderline of stability.

Once these prerequisites are satisfied, the approach proceeds as follows:
consider, as in the original formulation, two clusters of $N$ and $N'$
sites with $N'<N$. (Hereafter, primed and unprimed quantities will
refer to quantities pertaining to the small and large clusters respectively.)
The clusters {\it{must}} be chosen in such a way that the two different
sublattices of the given bipartite lattice are
mutually equivalent in both clusters.
Denoting the two sublattices by $A$ and $B$ and fixing the surrounding
sublattice magnetizations to the values $b'_1$ ($b_1$) and $b'_2$ ($b_2$)
on the $A$ and $B$ sublattices respectively, we then obtain by the usual
mean-field procedure expressions of the type $m'_{A(B)}(J',K',h',b'_1,b'_2)$
and $m_{A(B)}(J,K,h,b_1,b_2)$ for the sublattice magnetizations of the
small and big clusters. A mapping
$(J,K,h,b_1,b_2) \to (J',K',h',b'_1,b'_2)$ is now defined, in analogy with
eqs. \(map1) and \(scaleb1), through the requirements:
$$ m'_A(J',K',h',b'_1,b'_2) = {\s}^{d-y_h}m_A(J,K,h,b_1,b_2), \eqno(map2) $$
and
$$ b'_i={\s}^{d-y_h}b_i, \qquad i = 1,2.  \eqno(scalebi)$$
Equation \(scalebi) simply expresses the intuitively
appealing requirement that the boundary magnetizations scale the same
way as the bulk ones, as in the original MFRG.\refto{mfrg} We adopt
the usual definition ${\s}=(N/N')^{1/d}$ for the rescaling factor. Setting
again $h=h'=0$, expanding both sides of \(map2) to linear order in
$b'_{1,2}$ and $b_{1,2}$ and using \(scalebi) we
arrive at the flow equations for the coupling constants
$$ {{\partial m'_A(J',K',h',{\bf{b'}})}\over {\partial b'_1}}
   \bigg|_{h'=0,{\bf{b'}}=0} =
   {{\partial m_A(J,K,h,{\bf{b}})}\over {\partial b_1}}
	  \bigg|_{h=0,{\bf{b}}=0} \eqno(fl21) $$
and
$$ {{\partial m'_A(J',K',h',{\bf{b'}})}\over {\partial b'_2}}
   \bigg|_{h'=0,{\bf{b'}}=0} =
	  {{\partial m_A(J,K,h,{\bf{b}})}\over {\partial b_2}}
			\bigg|_{h=0,{\bf{b}}=0}, \eqno(fl22) $$
where we have used ${\bf{b'}}$ and ${\bf{b}}$ as a shorthand for
$(b'_1,b'_2)$ and $(b_1,b_2)$. Due to the equivalence of the two
sublattices, the individual subcluster magnetizations obey the symmetry
$m_A(J,K,h,b_1,b_2) = m_B(J,K,h,b_2,b_1)$ and the same for the
primed quantities,
thus the flows obtained if we use the $B$-sublattice magnetizations
to perform the mapping are identical to \(fl21) and \(fl22) above.
Hence, the method generates {\it{exactly two}} independent flow equations,
which are necessary and sufficient for the
renormalization of the two-parameter Hamiltionian.

After the fixed points ${\bf{J}}_* \equiv (J_*,K_*)$ of the flows
have been found, the nonmagnetic eigenvalues $\lambda_{1,2}$ and the
critical exponents $y_{1,2}=\ln \lambda_{1,2}/\ln {\s}$,
as well as the associated eigenvectors, are
evaluated by the usual procedure of linearizing around ${\bf{J}}_*$. The
remaining magnetic exponent is calculated through the relation
$$ {{\partial m'_A({\bf{J'}},h',{\bf{b'}})}\over{\partial h'}}
   \bigg|_{{\bf{J'}}={\bf{J}}_*,h'=0,{\bf{b'}}=0} =
   {\s}^{d-2y_h}
   {{\partial m_A({\bf{J}},h,{\bf{b}})}\over{\partial h}}
   \bigg|_{{\bf{J}}={\bf{J}}_*,h=0,{\bf{b}}=0}. \eqno(yh2) $$
Once more, due to the symmetry between $m_A$ and $m_B$ mentioned above,
it is irrelevant which of the two is used for the evaluation of $y_h$.
For the same reason, we can also work with the total magnetizations
$m'=m'_A+m'_B$ and $m=m_A+m_B$, and derive $y_h$ from the scaling
of the susceptibilty $\chi = {\partial m}/{\partial h}$ through
$$\chi' = {\s}^{d-2y_h}\chi, \eqno(chi)$$
which will be useful later. We also note that, although the feature
of working in the neighborhood of $b'=b=0$ is appropriate to the
study of second-order transitions, the flow equations presented
above are capable of producing also low-temperature
fixed points with the associated $y_h=d$ magnetic exponent which is
the signature of a first-order phase change.\refto{first}

Having developed the theoretical framework for our approximate
renormalization technique, we proceed in the following two Sections
with specific applications.

\head{\bf{III. Ising model with crossing bonds}}
\taghead{3.}

In this section we apply the extended MFRG (EMFRG) to the case of
the Ising model with first- and second-neighbor interactions
(Ising model with crossing bonds.) The Hamiltonian reads as
$$ H \equiv -\beta {\cal{H}} =
   h\sum_i s_i + J \sum_{<ij>}s_is_j + K \sum_{<<ij>>}s_is_j, \eqno(ising2) $$
where the second sum is carried over nearest sites and the third over
next nearest sites. The model satisfies requirement (i) of Section II;
in order to meet the requirement of a bipartite lattice, we must restrict
our choices: in two dimensions, we will consider the square ({\sq}) lattice,
which can be split into two square sublattices. Moreover, the model
on the square lattice has three types of ground states (for $h=0$): a doubly
degenerate ferromagnetic (FM) one in the region $\{J > 0; K > -J/2\}$,
a doubly degenerate antiferromagnetic (AFM) one in the region
$\{J < 0; K > J/2\}$ and a four-fold degenerate one with
super-antiferromagnetic (SAF) order in the remaining region
$ K<-|J|/2$.
The SAF-ground states consist of alternating
rows or columns of up and down spins. In order to satisfy requirement (iii)
of Section II, we shall {\it{only}} consider the flows in the subspace
$K \geq 0$ in which we have only the FM and AFM ground states
seperated by the borderline of stability $J=0$.

In three dimensions, we
will study the simple cubic ({\sc}) lattice which separates into two
face-centered cubic ({\fcc}) sublattices, and the body-centered cubic one
({\bcc})
which separates into two
simple cubic sublattices. Again, for reasons similar to
those of the two-dimensional case, we will only examine the flows in the
subspace $K \geq 0$; for both the {\sc}- and {\bcc}-models, the ground
states are again FM for $J>0$ and AFM for $J<0$.
We will present the calculation
in some detail for the {\sq}-model only, and just describe the
results for the {\sc}- and {\bcc}-cases, since the
essential characteristics of the flows are quite similar
for all three lattices.

{\it{A. Square lattice.}}
For the {\sq}-model, we take as the small cluster the $N'=2$ spins joined
by an elementary bond, and as the big cluster the $N=4$ spins around an
elementary plaquette of the lattice (fig. 1). These are the two smallest
clusters we are allowed to consider which satisfy the requirement of
equivalence of the sublattice-clusters. We take the points $A,C,\ldots$
forming the $A$-sublattice and the points $B,D,\ldots$ forming the
$B$-sublattice. With the boundary magnetizations fixed to the values
$b'_1$ ($b_1$) and $b'_2$ ($b_2$) for the $A$- and $B$-sublattices,
we obtain the effective Hamiltonians:
$$ H'(s_A,s_B) = J'\bigl[s_As_B+3(s_Ab_2+s_Bb_1)\bigr]+
				 4K'(s_Ab_1+s_Bb_2)+h'(s_A+s_B), \eqno(effis2)
$$
for the $N'=2$-cluster and
$$ H(s_A,s_B,s_C,s_D) =
   J\biggl[s_As_B+s_Bs_C+s_Cs_D+s_Ds_A+2[(s_A+s_C)b_2+(s_B+s_D)b_1]\biggr]+ $$
$$ K\biggl[s_As_C+s_Bs_D+3[(s_A+s_C)b_1+(s_B+s_D)b_2]\biggr]+
   h(s_A+s_B+s_C+s_D) \eqno(effis4) $$
for the $N=4$-cluster.
Using eqs. \(effis2) and \(effis4) above, we find the sublattice
magnetizations per site as
$$ m'_A = {{\sinh[(3J'+4K')(b'_1+b'_2)+2h']-
			e^{-2J'}\sinh[(3J'-4K')(b'_1-b'_2)]}\over
           {\cosh[(3J'+4K')(b'_1+b'_2)+2h']+
			e^{-2J'}\cosh[(3J'-4K')(b'_1-b'_2)]}} \eqno(mis2) $$
and
$$ m_A = \Bigl\{e^{4J+2K}\sinh\biggl[(4J+6K)(b_1+b_2)+4h\biggr]+
		 2\sinh(4Jb_2+6Kb_1+2h)- $$
$$  e^{-4J+2K}\sinh\biggl[(4J-6K)(b_1-b_2)\biggr]\Bigr\}\times$$
$$ \Bigl\{e^{4J+2K}\cosh\biggl[(4J+6K)(b_1+b_2)+4h\biggr]+
   2\cosh(4Jb_2+6Kb_1+2h)+$$
$$ 2\cosh(4Jb_1+6Kb_2+2h)+
 e^{-4J+2K}\cosh\biggl[(4J-6K)(b_1-b_2)\biggr]+
   2e^{-2K}\Bigr\}^{-1}. \eqno(mis4) $$
Combining \(fl21) and \(fl22) with \(mis2) and \(mis4) above we
obtain the $(J,K) \to (J',K')$ flows in the form
$$ {{3J'+4K'-e^{-2J'}(3J'-4K')}\over{1+e^{-2J'}}}=
   {{4J+6K+12Ke^{-4J-2K}-e^{-8J}(4J-6K)}\over
	{1+4e^{-4J-2K}+e^{-8J}+2e^{-4J-4K}}} \eqno(flis1) $$
and
$$ {{3J'+4K'+e^{-2J'}(3J'-4K')}\over{1+e^{-2J'}}}=
   {{4J+6K+8Je^{-4J-2K}+e^{-8J}(4J-6K)}\over
	   {1+4e^{-4J-2K}+e^{-8J}+2e^{-4J-4K}}}. \eqno(flis2) $$
In the $K\geq 0$-subspace, the flows \(flis1) and \(flis2) have the
fixed points $(J_*,K_*)$ presented below.
\sk
\ni
(i) A stable low-temperature
fixed point $L_1=(+\infty,+\infty)$ corresponding to the
FM ground state.
\sk
\ni
(ii) A stable low-temperature fixed point $L_2=(-\infty,+\infty)$
corresponding to the AFM ground state.
\sk
\ni
(iii) A mixed low-temperature fixed point $L_3=(0,+\infty)$ which attracts
along the $J=0$ direction only and repels in all other directions. This
fixed point corresponds to the four degenerate FM and AFM configurations
that are the ground states of the model which, for $J=0$, reduces to
two decoupled {\sq}-Ising models with nearest-neighbor coupling $K$.
\sk
\ni
(iv) A critical point $C_1=(0.1590,0.1499)$ which represents the
ferromagnetic Ising criticality.
\sk
\ni
(v) A critical point $C_2=(-0.1590,0.1499)$ which represents the
antiferromagnetic Ising criticality.
\sk
\ni
(vi) A multicritical point $C_3=(0,0.3465)$ which represents
the criticality of the uncoupled {\sq}-Ising models described above, and
\sk
\ni
(vii) the stable, high-temperature fixed point $P=(0,0)$ representing the
paramagnetic phase.
\sk
\ni

The flows in the $K \geq 0$ subspace are shown in fig. 2. We note that some of
the flows emerge from two unstable fixed points located at $J=-2K,\;
J \to +\infty$ and $J=2K, J \to -\infty$,
located at the borderlines of stability of the
ground states, which are not shown. As a first
remark, we note that the flows are very similar to those obtained in the
past by the use of other RG techniques,\refto{{nauenberg},{vanleeuwen}}
displaying reflection symmetry about the $J=0$ axis and a cusp of
the critical surface at the point $C_3$.
However, the RG flows are obtained here in a much simpler way which
is also readily generalizable to three-dimensional models in a very
straigthforward fashion. The flows are also similar to those obtained
in earlier work by Plascak,\refto{plas} but with the significant
difference that here we avoid the ambiguities in the flows along the
$J=0$ axis found in that work.
In particular, we are capable to obtain the
multicritical fixed point $C_3$ {\it{on the $J=0$ axis}} which was
missing in that approach and on which, as we demonstrate below, certain
important relations between the critical exponents are satisfied to a
reasonable degree of accuracy. On the other hand, an obvious
deficiency of our approach in its present formulation is that the
theory is unable to locate the SAF-critical fixed point at $J=0,\;K_{*}<0$
since the way of separating the sublattices makes it inapplicable
at present to the region in which the SAF-fixed point lies.

For the calculation of the magnetic exponent $y_h$, a special treatment
is required for the AFM fixed points. As can be seen from Eq. \(chi),
when $h$ is relevant, {\it{i.e.}} $y_h>0$ (and $y_h>d/2$ of course) we
have $\chi' < \chi$. On the other hand, when $h$ is irrelevant and thus
$y_h<0$ we obtain the inequality $\chi'>\chi$. However, the last
inequality itself only implies $y_h<d/2$, and not necessarily $y_h<0$.
Indeed, the sign of $y_h$ is very sensitive to the choice of the
rescaling factor ${\s}$. In order to avoid complications related
to the choice of this factor, and keep the discussion as simple as
possible at present, we will characterize the magnetic field as
irrelevant whenever we obtain $\chi'>\chi$, and this is indeed what
we get at the AFM-criticality fixed point $C_2$.

For the low-temperature fixed points $L_1$ and $L_3$ we obtain the correct
magnetic exponent $y_h=d$, a manifestation of a first-order phase
transition.\refto{first} For the point $L_2$ the magnetic field is irrelevant,
whereas at the paramagnetic fixed point $P$ we find correctly the
result $y_h=d/2$. The locations of the $T>0$ critical
points and the corresponding critical exponents are summarized in
Table I.

The critical lines intersect the $K=0$ axis at the points $J_c=\pm 0.336$,
which is the estimate from our theory for the
FM and AFM critical couplings of the
nearest-neighbor Ising model, to be compared with the exact result\refto{domb}
$J_c=\pm 0.441$. This result is of the same quantitative accuracy with
that obtained from the original MFRG with a $2 \to 1$ mapping, namely
$J_c=0.346$; the latter is also the value of $K_*\equiv K_c$ at the
fixed point $C_3$ which represents the FM criticality of the uncoupled
{\sq}-Ising models with nearest neighbor interactions {\it{only}}.
The thermal exponent $y_1=0.760$ (at points $C_1$ and $C_2$) is in
reasonable agreement with the exact result $y_t=1$, and the magnetic
exponent $y_h=1.487$ differs from the exact one $y_h=15/8$ by about
$20\%$.

At the multicritical fixed point $C_3$ we have an exponent $y_1=1.073$
with the corresponding eigenvector along the $(1,0)$-direction and an
exponent $y_2=0.600$ with the corresponding eigenvector along the
$(0,1)$-direction. This allows us to identify the latter with the
thermal exponent $y_t$ of the uncoupled Ising lattices, and we call
$y_1=y_s$ to comply with the terminology introduced in \ref{vanleeuwen}.
As a first remark, we note that $y_t$ differs from the thermal exponent
$y_1$ for the points $C_1$ and $C_2$ but this is not surprising in view
of the fact that our method is a $4 \to 2$ mapping in general, but for
$J=0$ it reduces effectively to a $2 \to 1$ mapping; different degrees of
approximation in the mappings yield different estimates for the
critical exponents, as expected. On the other hand, the magnetic exponent
$y_h=1.415$ at $C_3$ is not too different form $y_h=1.487$ at $C_1$ and
$C_2$. According to van Leeuwen\refto{vanleeuwen} the exponents
$y_s$ and $y_t$ must satisfy the relations:
$$ y_s = 2y_h-d \eqno(rule1) $$
and
$$ y_s/y_t = \gamma_{\rm{Ising}}. \eqno(rule2) $$
Using the results in Table I, we find that the rhs of \(rule1) is equal to
0.830, whereas the lhs is 1.073, which shows once more that the critical
exponents are not evaluated very accurately in the MFRG, at least with
small clusters and with the naive definition of ${\s}$; however, the ratio
$y_s/y_t$ is equal to 1.789, and the exact result for the rhs of
\(rule2) is
$\gamma_{\rm{Ising}} = 1.75$, which shows that {\it{ratios}} of
the critical exponents come out rather accurately in this approach, since
they are independent of the precise definition of the rescaling factor.

We now extend the approach to three dimensions, studying the flows for
the simple cubic and body-centered cubic models, always in the subspace
$K \geq 0$.

{\it{B. Simple cubic lattice.}} For the {\sc}-lattice the two clusters
are again the elementary bond ($N'=2$) and the elementary plaquette
($N=4$), once more the smallest possible clusters which satisfy the
requirements of the approach. The low- and high temperature fixed
points $L_1$, $L_2$, $L_3$ and $P$ are identical to the {\sq}-case,
also with the correct magnetic exponents. Once more, we find the
two critical points $C_1$ and $C_2$ and the multicritical point
$C_3$ whose coordinates and critical exponents are summarized in
Table II. The overall flow pattern is identical to the {\sq}-case.

The critical lines now intersect the $K=0$-axis at the points $J_c=\pm 0.192$
to be compared with the `exact' result\refto{domb} $J_c=\pm 0.222$.
At the critical points $C_1$ and $C_2$ we find the thermal exponent
$y_1=0.506$ which differs quite a bit from the `exact' result\refto{legui}
$y_t=1.587$; the magnetic exponent $y_h=1.776$ is somewhat closer to
the value\refto{legui} $y_h=2.485$. The value $K_*$ of the point
$C_3$ gives an estimate of the critical coupling of the
{\fcc} nearest-neighbor Ising model, since for $J=0$ the model decouples
into the two independent {\fcc} sublattices of the {\sc}-lattice.
We obtain $K_c \equiv K_* = 0.091$ to be compared with the
estimate\refto{domb}
$K_c=0.102$.
We note in passing
that the MFRG inherits from the underlying mean-field ideas the
charactersistic that the values of the critical couplings improve,
for the same degree of approximation, as the coordination number of the
lattice increases. The value of the thermal exponent, however, does not:
at the point $C_3$, $y_t\equiv y_2=0.344$ which is worse than the
value at $C_1$ and $C_2$;
ideally, all these estimates should converge to the exact 3D-Ising values
as one considers larger and larger clusters. The magnetic exponent at
$C_3$, $y_h=1.688$, is again not too different from its value at $C_1$.
Regarding the relation \(rule1) we obtain 0.443 for the lhs and 0.377
for the rhs, whereas the ratio $y_s/y_t = 1.288$ is not too far away from
the `exact' value $\gamma_{\rm{Ising}}=1.238$ (\ref{mcrg})
required from eq. \(rule2).
Again, the ratio of the critical exponents
is very accurate, although the individual exponents are not.

{\it{C. Body-centered cubic lattice.}} Here, the small cluster is chosen
to be the nearest-neighbor bond ($N'=2$). For the large cluster, there
are two $N=4$ choices that satisfy the requirements:
the $ABCD$-tetrahedron and the $ADCE$ parallelogram shown in fig. 3.
Both mappings yield identical low- and high-temperature fixed
points, with the correct magnetic exponents. Moreover, they yield
the same multicritical point $C_3=(0,0.203)$ but they differ slightly in the
nonmagnetic critical exponents at $C_3$ as well as at the locations
and exponents of the critical points $C_1$ and $C_2$. The flow patterns
are again very similar to those presented above for the other lattices.
Although the critical fixed points are different, the
{\it{critical lines}} from the two mappings are almost identical.
This demonstrates that the method is robust (insensitive) to the choice
of the large cluster.

The results are summarized in Tables III(a) and III(b). The value
$K_*=0.203$ at the fixed pint $C_3$
is an estimate for the critical coupling of the {\sc}-Ising
model with nearest neighbor coupling, again not too far from the exact
value, 0.222, and the previous {\sc} estimate, 0.192. The critical lines
now intersect the $K=0$-axis at the points $J_c = \pm 0.140$ for the
tetrahedron $\to$ bond mapping and
$J_c=\pm 0.139$ for the parallelogram $\to$ bond mapping
which are the estimates for the {\bcc}-Ising critical coupling,
to be compared with the best estimate\refto{domb} $J_c = \pm 0.157$.
A comparison of the critical exponents with the exact ones shows that
the tetrahedron $\to$ bond mapping yields better results not only in
terms of their numerical values, but also in the sense that it predicts
almost identical values for the thermal and magnetic exponents for
the points $C_1$ and $C_3$, as required by universality.
The relations \(rule1) and \(rule2) are satisfied within
an error of at most $10\%$.

The above discussion demonstrates that the RG-method we propose here
yields very satisfactory results for some standard `test'-models, and
thus it is a plausible technique for the renormalization of more
complicated Hamiltonians.

\head{\bf{IV. Morphological model}}
\taghead{4.}

We consider in this Section a
phenomenological morphological Hamiltonian on a {\bcc}-lattice,
introduced recently in order to model the phase behavior of
microemulsions.\refto{wagner} Here, we outline the basic ideas in the
derivation of the model, and refer the reader to \ref{wagner} for details.
Let us consider a three-dimensional Bravais lattice with periodic
boundary conditions, having $N$ sites and volume $V$, whose Wigner-Seitz
(WS) unit cells contain either bulk water or oil.
In addition, the system
contains amphiphilic molecules which are supposed to form an incompressible
membrane in the interface between oil and water, defining in this way
a Gibbs dividing surface between the two bulk phases. After choosing an
orientation for this interface, a collection of water cells fixes
uniquely an interfacial pattern, and vice versa.
We denote a cell to be `occupied' if it contains water, and
`empty' if it contains oil.
The morphological features of each pattern can be
described by the Minkowski functionals;
in three dimensions, these are the
geometric invariants: covered volume, surface area, integral
mean curvature of the interface, and the combinatorial Euler
characteristic $X$ of the pattern. We have $X=$ (number of disconnected
components)-(number of handles)+(number of cavities) in three dimensions.
In the present model the interfaces have no holes.
The family of the Minkowski functionals
is characterized by a theorem which asserts that any real-valued,
additive, motion-invariant and continuous functional defined on the
collection of the $2^N$ configurations is a
linear combination of the Minkowski functionals.\refto{{hadwiger},{weil}}
In order to deal with the statistical morphology of the interfacial
membrane, we therefore take the Hamiltonian to be of the generic form:
$$ {\cal{H}}=\sum_{\alpha=0}^{3}h_{\alpha}V_{\alpha}, \eqno(mink) $$
where $h_{\alpha}$ are energy parameters and $V_{\alpha}$ are,
within proportionality factors, the dimensionless
Minkowski functionals as follows: denoting by ${\cal{V}}$, ${\cal{A}}$,
${\cal{M}}$ and $X$ the covered volume, exposed area, integral mean
curvature and Euler characteristic of the pattern formed by the
full cells, we have $V_0={\cal{V}}/(8\sqrt{2}l^3)$,
$V_1={\cal{A}}/[3(4\sqrt{2}+2)l^2]$, $V_2={\cal{M}}/(6\pi l)$ and
$V_3=X$. The various factors arise from the definition of the Minkowski
functionals via Steiner's formula, and their values for the {\bcc} WS
polyhedron (see the Appendix and Table I of \ref{wagner}.)
The length scale $l$ is the edge
length of the WS-unit cell of the {\bcc}-lattice, $l=a\sqrt{2}/4$, where
$a$ is the {\bcc} lattice constant (see fig. 3.)

Using the property of additivity, one can derive concise expressions for
the Minkowski functionals in terms of the occupation numbers
$u_i=0,1$ ($u_i=0$ if the cell is occupied, $u_i=1$ if it is empty)
and in \ref{wagner} the Hamiltonian was written down explicitly in that
representation.
Here, it will be more useful to write the Hamiltonian in terms of
`Ising spin' variables, $s_i=\pm1$ with $s_i=1$ denoting a full site,
and $s_i=-1$ an empty one ({\it{i.e.}} $s_i=1-2 u_i$.)
Setting ${\bf{s}}=(s_1,s_2,\ldots,s_N)$, the final expression
reads as
$$ {\cal{H}}({\bf{s}}) = \sum_{\alpha=0}^{3}h_{\alpha}V_{\alpha}({\bf{s}}) =
h_0\Biggl[{N\over{2}}+{1\over{2}}\sum_{i} s_i\Biggr]+$$
$$ {{h_1}\over{4}}\Biggl[N-{{\sqrt{3}}\over{4\sqrt{3}+2}}
   \sum\nolimits^{'}s_is_j-
{{2}\over{3(4\sqrt{3}+2)}}\sum\nolimits^{''}s_is_j\Biggr]+$$
$$ {{h_2}\over{4}}\Biggl[-\sum_{i} s_i + {1\over{12}}
   \sum\nolimits^{'''} s_is_js_k\Biggr]+
 {{h_3}\over{8}}\Biggl[-N+\sum\nolimits^{'}s_is_j-
	{1\over{2}}\sum\nolimits^{''''}s_is_js_ks_l\Biggr]. \eqno(hamiltonian)
$$
The primed and double-primed sums are carried
over nearest and next-nearest neighbor bonds, respectively. The
triple-primed sum runs over isosceles triangles, two of the sides of
the triangles being first neighbor bonds and the third being a second
neighbor bond. Such are the $ABC$, $BCD$, $ABD$, and $ACD$ triangles
in fig. 3, for instance. Finally, the four-primed sum runs over tetrahedra
whose faces are isosceles triangles as above, e.g. the $ABCD$-tetrahedron
of fig. 3.
Under the interchange $s_i \to -s_i$, the integral mean curvature
is odd, whereas the exposed area and Euler characteristic
are even.

We are going to deal
exclusively with the case $h_2=0$ (no spontaneous internal curvature.)
For $h_0=h_2=0$, the ground states (GS's) of the model are the following:
a doubly degenerate ferromagnetic GS (oil or water, OW) in the region
$$ \Biggl\{h_3 < {\sqrt{3}\over{2\sqrt{3}+1}} h_1 ;
   h_3 > -h_1\Biggr\}; \eqno(ow) $$
a doubly degenerate antiferromagnetic GS (`plumber's nightmare', PN) in
the region
$$ \Biggl\{h_3 > {\sqrt{3}\over{2\sqrt{3}+1}} h_1 ;
   h_3 > {{2\sqrt{3}-1}\over{5(2\sqrt{3}+1)}} h_1 \Biggr\}; \eqno(pn) $$
and a four-fold degenerate `droplet'-phase in the region
$$ \Biggl\{h_3 < -h_1 ;
   h_3 < {{2\sqrt{3}-1}\over{5(2\sqrt{3}+1})} h_1\Biggr\}. \eqno(droplet) $$
The `droplet' phase is realized when one of the two {\sc}-sublattices has
ferromagnetic and the other antiferromagnetic order. The subspace
$h_3>0$ is covered completely by the OW and PN-ground states,
separated by the borderline of stability $h_1=(2\sqrt{3}+1)h_3/\sqrt{3}$;
thus, according
to the general requirements laid out in Section II, we are going to
consider the flows in the subspace $h_3 \geq 0$ only.\refto{remark}

Dropping the uninteresting spin-independent constants from the
Hamiltonian, and defining:
$$ h = -{{\beta h_0}\over{2}}{\rm{;}}\;\;
J = {{\beta h_1}\over{24(4\sqrt{3}+2)}}{\rm{;}}\;\;
K = {{\beta h_3}\over{16}}, \eqno(redef) $$
we arrive at the expression
$$ H \equiv -\beta{\cal{H}} =
   h\sum_{i} s_i + (6\sqrt{3}J-2K)\sum\nolimits^{'} s_i s_j +
   4J\sum\nolimits^{''} s_is_j +
   K\sum\nolimits^{''''} s_is_js_ks_l. \eqno(boltz) $$
For the choice $h_3=0$ the model reduces to a conventional {\bcc} Ising
model with an additional second-neighbor coupling, namely
$$ H = h\sum_i s_i + L \sum\nolimits^{'} s_is_j +
	   \alpha L \sum\nolimits^{''} s_is_j, \eqno(special) $$
where $L=\beta \sqrt{3}h_1/(16\sqrt{3}+8)$ and $\alpha=2/(3\sqrt{3})$
is the ratio of second
to first neighbor coupling.

We are now interested in the renormalization of Hamiltonian \(hamiltonian).
When $h_0=h_2=0$ a real-space RG will generate flows with at least three
parameters corresponding to the three even interactions in \(boltz),
whereas the initial morphological Hamiltonian \(mink) contains only
two, $h_1$ and $h_3$ (or $J$ and $K$ in \(redef), \(boltz)) in the even
subspace. However, after coarse-graining the effective Hamiltonian must
still be additive, motion-invariant and continuous as the initial one.
Thus, it must be again a linear combination of the Minkowski functionals,
{\it{i.e.}} of the form \(mink). Our extension of the MFRG allows us to
impose this requirement quite naturally, by assuming that the
renormalized Hamiltonian for the $N'$-cluster always has the form \(boltz).

Clearly, the model satisfies all prerequisites of Section II, and we can
proceed with the EMFRG approach.
We choose for the small cluster the $AB$-bond ($N'=2$) and for the large one
the $ABCD$-tetrahedron ($N=4$) shown in fig. 3. The points $A,D,\ldots$
form the $A$-sublattice and the points $B,C,\ldots$ form the $B$-sublattice.
Some details on the derivation of the flows are given in the Appendix.
The flow equations read as
$$ {{(42\sqrt{3}J'-14K')\sinh(6\sqrt{3}J'-2K')+
	 24J'\cosh(6\sqrt{3}J'-2K')}\over
    {\cosh(6\sqrt{3}J'-2K')}} = $$
$$ =\Biggl\{e^{(24\sqrt{3}+8)J-7K}\bigl[(72\sqrt{3}+40)J-22K\bigr]+
	2 e^{-K}(40J-2K)- $$
$$ e^{(-24\sqrt{3}+8)J+9K}\bigl[(72\sqrt{3}-40)J-26K\bigr]\Biggr\}\times $$
$$ \Biggl\{e^{(24\sqrt{3}+8)J-7K}+4e^{-K}+2e^{-8J+K}+
		   e^{(-24\sqrt{3}+8)J+9K}\Biggr\}^{-1} \eqno(mfl1) $$
and
$$ {{(42\sqrt{3}J'-14K')\cosh(6\sqrt{3}J'-2K')+
      24J'\sinh(6\sqrt{3}J'-2K')}\over
    {\cosh(6\sqrt{3}J'-2K')}} = $$
$$ =\Biggl\{e^{(24\sqrt{3}+8)J-7K}\bigl[(72\sqrt{3}+40)J-22K\bigr]+
    2 e^{-K}(72\sqrt{3}J-24K)+ $$
$$ e^{(-24\sqrt{3}+8)J+9K}\bigl[(72\sqrt{3}-40)J-26K\bigr]\Biggr\}\times $$
$$ \Biggl\{e^{(24\sqrt{3}+8)J-7K}+4e^{-K}+2e^{-8J+K}+
           e^{(-24\sqrt{3}+8)J+9K}\Biggr\}^{-1}. \eqno(mfl2) $$
The above flows have
in the subspace $K \geq 0$ the fixed points $(J_*,K_*)$ listed below.
\sk
\ni
(i) A stable low-temperature fixed point $L_1=(+\infty,+\infty)$ representing
the ferromagnetic (OW) ground state.
\sk
\ni
(ii) A stable low-temperature fixed point $L_2=(-\infty,+\infty)$ representing
the antiferromagnetic (PN) ground state.
\sk
\ni
(iii) A mixed low-temperature fixed point $L_3=(J_*,3\sqrt{3}J_*)$ with
$J_* \to +\infty$. This point attracts in the direction $K=3\sqrt{3}J$
({\it{i.e.}} $h_3=\sqrt{3}h_1/(2\sqrt{3}+1)$) and repels in the other
directions. It corresponds to the four-fold degenerate ground states
of the model when the couplings $h_1$ and $h_3$ have the ratio given
above, and in which case the first-neighbor coupling vanishes; then,
the OW and PN phases are all degenerate at $T=0$.
\sk
\ni
(iv) A critical point $C_1=(1.656\times 10^{-2}, 5.306\times 10^{-2})$
which represents the Ising ferromagnetic criticality.
\sk
\ni
(v) A critical point $C_2=(1.407\times 10^{-2}, 9.858\times 10^{-2})$
which represents the Ising antiferromagnetic criticality.
\sk
\ni
(vi) A multicritical point $C_3=(2.785\times 10^{-2}, 1.447\times 10^{-1})$
and
\sk
\ni
(vii) the high-temperature, stable fixed point $P=(0,0)$ representing the
paramagnet.

The flows are shown in fig. 4(a); due to the particular representation
chosen to show the flows, the fixed points $L_1$ and $L_3$ coincide in
this figure. We obtain the correct magnetic exponent
$y_h=d$ at $L_1$ and $L_3$ and $y_h=d/2$ at $P$.
The subspace $h_3=\sqrt{3}h_1/(2\sqrt{3}+1)$ is an
eigenspace of the flows,
with the fixed point $C_3$ lying in this subspace.
Flows that start in this space remain in it,
running towards the point $L_3$ if they start above $C_3$ or towards
the point $P$ if they start below $C_3$. Thus, in the
special case of a model with vanishing first-neigbor coupling, the
flows maintain that property.
The parameter space is thus separated into three
basins of attraction (see fig. 4(a)). Flows starting in
the region enclosed in the
`triangle' formed by the two critical lines and the axis $h_3=0$ run
towards the $P$-point. Flows starting outside the triangle and on the
right of the line $P-C_3-L_3$ run to the OW fixed point $L_1$, always
remaining in the region $h_3 < \sqrt{3}h_1/(2\sqrt{3}+1)$; and flows
that start outside the triangle but on the left of the line
$P-C_3-L_3$ run initially close to $L_3$, but eventually they turn around
to end up in the PN fixed point $L_2$, staying always in the
subspace $h_3 > \sqrt{3}h_1/(2\sqrt{3}+1)$.

The locations of the $T>0$-fixed points and the associated critical
exponents are summarized in Table IV. We note that the method
predicts identical values for the thermal exponent $y_1$ for both
ferromagnetic and antiferromagnetic Ising critical points,
as should. The value $y_1=0.691$  is comparable to the value
$y_1=0.653$ obtained for the {\bcc} Ising model with crossing bonds
using the same kind of mapping (tetrahedron $\to$ bond) discussed in
Section III (see Table III(a).) As further evidence for the consistency
of the method, we note that the critical line of the $C_1$ fixed point
intersects the $h_3=0$-axis at the point $\beta h^{c}_1=2.019$ (see
fig. 4(b)). Accordingly,
the critical coupling of the Hamiltonian \(special)
(to which the model reduces in the case $h_3=0$) is predicted
to be $L_c=0.098$. On the other hand, the line corresponding to
$\alpha=2/(3\sqrt{3})$ intersects
the critical line of the ferromagnetic fixed point $C_1$ of the
{\bcc} Hamiltonian studied in Section III at a point whose abscissa is
equal to $0.105$. The two estimates differ by less than $7\%$.

In order to obtain more detailed information about the phase behavior
of the model, we have to invoke the results from other techniques as well,
for example from the simple mean-field approximation. The reason is that
the proposed RG scheme has certain limitations due to the feature of
always expanding the cluster magnetization around vanishingly small
expectation values $\langle s_i \rangle$ of the surrounding spins.
Consequently,
not all of the points lying on the `critical lines' correspond to
true criticalities. Let us consider, for instance, the
mean-field expression for the free energy per site
which reads as\refto{wagner}
$$ \beta f(m) = {{\beta h_1}\over{4}}(1-m^2)+
				{{\beta h_3}\over{8}}(-1+4m^2-3m^4)+ $$
$$  {{1+m}\over{2}}\ln\biggl({{1+m}\over{2}}\biggr)
   +{{1-m}\over{2}}\ln\biggl({{1-m}\over{2}}\biggr), \eqno(fmfa) $$
where $m = \langle s_i \rangle$. Expanding the free energy
about $m=0$ up to $O(m^4)$ and dropping the uninteresting constants
we obtain
$$ \beta f(m) =
  \biggl(-{{\beta h_1}\over{4}}+{{\beta h_3}\over{2}}+{1\over{2}}\biggr) m^2+
  \biggl(-{{3\beta h_3}\over{8}}+{1\over{12}}\biggr) m^4.
  \eqno(expand) $$
The requirement of vanishing coefficient of the quadratic term identifies
the line of diverging susceptibility, $\chi^{-1}=0$, and reads as
$$ \beta h_3 = {{\beta h_1-2}\over{2}}. \eqno(spinod) $$
However, not all the points defined by \(spinod) correspond to true
criticalities which happen only
when the coefficient of the $m^4$-term is positive.
This requirement yields a tricritical  point
$(\beta h^{tr}_1,\beta h^{tr}_3) = (22/9, 2/9)$. In fig. 4(b) we plot
the line \(spinod) along with the flows
and indicate the tricritical point by a dot. We note
that below the dot the attractive line of the fixed point
$C_1$ almost coincides with \(spinod). The true tricrical point of the
model must occur at a temperature lower than the mean-field prediction,
of course, but
taking for now the estimate for tricriticality
from the mean-field approximation for granted, we can give to the
attractive line of the point $C_1$ the following
interpretation: the segment of the
line below the dot corresponds to true
criticality. However, the part of the solid line above the dot
does not represent a line of critical points, but rather a line of
diverging susceptibility at $m=0$ which, however, does not correspond to
true criticalities because these are preempted the phase coexistence
between oil-rich and water-rich phases.\refto{remark2}
The inability
of the present RG scheme to distinguish between a critical line
and a line of preempted
criticalities lies in the expansions of the magnetizations about
vansihingly small values of the boundary spins, {\it{i.e.}} in the
assumption that the system is translationally invariant and has a vanishingly
small bulk magnetization. Similar considerations also apply to the
$C_2$-`critcal line'.

Another feature of the model found in the mean-field approximation and
confirmed by extensive Monte-Carlo simulations\refto{wagner} is the
possibilty of {\it{three phase coexistence}} between oil-rich,
water-rich and a middle disordered phase, characterized as a microemulsion
in view of its morphological (negative Euler characteristic) and
structural (a peak of the structure factor at nonzero wavevector)
properties. The three phase coexistence is caused by the competition
between a positive surface tension $\beta h_1 \sim 1$ of the incompressible
surfactant film, which tends to minimize the exposed area, and a positive
topological potential $\beta h_3 \sim 1$ which encourages structures
with a large surface area in order to accomodate many handles, and thus
it assists the entropic tendency to disperse the amphiphiles.
The microemulsion has a finite correlation
length $\xi$ of the order of a few lattice constants, {\it{i.e.}} it displays
short-range order (unlike the completely random mixture or `paramagnet'
in the magnetic language.) Clearly, since $\xi$ is neither vanishing
nor diverging, the microemulsion is not represented by a fixed point
in the flow diagram. We must, therefore, resort once more to external
information in order to identify the region in the flow diagram which
represents stable microemulsion phases. According to earlier
work,\refto{wagner} the middle phase occurs in a range of temperatures
$ 0.80 {< \atop \sim} k_{B}T/h_3 \leq 9/2 $, the upper limit being the
tricritical temperature, above which three-phase coexistence ceases to exist.
On the other hand, below the lower limit the middle phase again disappears
because it is replaced by a slightly disordered `plumber's nightmare'
phase. We take the combinations of the energy coefficients that give
a three-phase coexistence as `pointers' that indicate the region on the
flow diagram where the microemulsion is stable. In fig. 4(b) we mark
a few of those points, which are once more obtained form the
mean-field approximation; the latter has been found to be relatively
accurate in its predictions for the triple points, when compared to
the simulation results.\refto{wagner} Referring to this figure, we can
now assert that the neighborhood of the filled triangles below the $PC_3$
line, to the left of the solid line and above the tricritical point
roughly defines the domain of thermodynamic stability of the
microemulsion. Clearly, upon coarse-graining all points in the microemulsion
regime flow towards the paramagnet. In order to delimit the region of
stability of the microemulsion more accurately one could, for example,
study the small wavenumber behavior of the Fourier transform of the
correlation function within the RG scheme. However,
such a calculation lies beyond the
scope of this work.

\head{\bf{V. SUMMARY AND CONCLUSIONS}}
\taghead{5.}

We have presented a generalization of the mean-field renormalization
group method which provides a way of renormalizing Hamiltonians with
two coupling constants in the even interaction part. The method builds
on ideas which are similar to those of the original MFRG,
but it goes beyond the limitation of a single flow equation by employing
scaling assumptions about suitably chosen sublattice magnetizations.
We applied the method to several model Hamiltonians, obtaining
satisfactory results for the flow diagrams. An obvious challenge for the
future is the development of these ideas even further, so that we will
be able to deal with Hamiltonians involving more than two parameters.
A possible way to achieve this goal will be through a separation
of the lattice into
more than just two suitably chosen sublattices; however, since the EMFRG
always yields as many flow equations as sublattices, it may be necessary
in some cases to augment the original Hamiltonian by a suitably chosen
number of interactions and coupling constants until the number of
couplings matches the number of sublattices. Then, the flows would be
obtained in this enlarged parameter space, and one could look at the
flows in the original, restricted domain by taking the appropriate
`cuts' in Hamiltonian space.

The morphological model of Section IV includes all additive geometrical
invariants whose thermal averages are extensive. The manifest additivity
of the Hamiltonian is a sufficient, but not necessary condition for the
thermodynamic requirement of extensitivity of the internal energy.
Therefore, one should not exclude {\it{a priori}} terms in the Hamiltonian
which are not additive; in particular, a non-additive `curvature-square'
term, which is employed in most current models of microemulsions, is
missing from the Hamiltonian. In the original paper,\refto{wagner}
it was argued that
the model deals with length scales exceeding the
persistence length\refto{degennes} $\xi_{\kappa}$, where the above
contribution (also called `bending energy') can be omitted because
the scale-dependent bending rigidity has been renormalized away.
A non-perturbative renormalization of a Hamiltonian including
the surface area and bending energy terms only was presented
recently;\refto{mecke} our model is complementary to that of \ref{mecke}
in that it includes the Euler characteristic term, but not the
bending rigidity.
It would be desirable, therefore, to start with a Hamiltonian that
includes the Minkowski functionals {\it{and}} the bending energy and
proceed with its renormalization, in order to see the crossover from
the rigidity-dominated regime (for lengths scales below $\xi_{\kappa}$) to
the regime above $\xi_{\kappa}$ where the thermal fluctuations dominate
over the rigidity and the membrane is crumpled. We plan to return to
this problem in the future.
\bigskip
\noindent
\centerline{\bf{ACKNOWLEDGMENTS}}
\medskip
\noindent
We wish to thank Professor H. Wagner for suggesting this problem to one
of us (C.N.L.) and for helpful discussions.
This work was started while C.N.L. was a postdoctoral fellow at the
Sektion Physik der
Ludwig-Maximilians-Universit\"at M\"unchen. C.N.L.
wishes to thank the Alexander von Humboldt-Stiftung
for a fellowship that supported him throughout his stay in Munich,
and also acknowledges the
support by the Human Capital and Mobility Programme
of the Commission of the European Communities, Contract No
ERBCHBICT940940, during his stay in Trieste.

\head{\bf{Appendix}}
\taghead{A.}

Here we outline the steps for the derivation of the flow equations for
the morphological Hamiltonian of Section IV, eq. \(boltz). Consider
first the small cluster $AB$, $N'=2$. Fixing all the surrounding
$A(B)$-sublattice spins to the value $b'_1(b'_2)$, and using \(boltz)
and figure 3, we arrive at the effective Hamiltonian of the $N'=2$
cluster of the form:
$$H'(s_A,s_B)=h'S'_0+(6\sqrt{3}J'-2K')S'_1+4J'S'_2+K'S'_4, \eqno(boltz2) $$
where
$$ S'_0=s_A+s_B, \eqno(sp0) $$
$$ S'_1=s_As_B+7(s_Ab'_2+s_Bb'_1), \eqno(sp1) $$
$$ S'_2=6(s_Ab'_1+s_Bb'_2) \eqno(sp2) $$
and
$$ S'_4=6b'_1b'_2[s_As_B+3(s_Ab'_2+s_Bb'_1)]. \eqno(sp4) $$
Simliarly, the effective Hamiltonian for the $ABCD$-cluster $(N=4)$
has the form
$$H(s_A,s_B,s_C,s_D)=hS_0+(6\sqrt{3}J-2K)S_1+4JS_2+KS_4, \eqno(boltz4) $$
where
$$ S_0=s_A+s_B+s_C+s_D, \eqno(s0) $$
$$ S_1=s_As_B+s_Bs_D+s_Ds_C+s_Cs_A+6[(s_A+s_D)b_2+(s_B+s_C)b_1], \eqno(s1) $$
$$ S_2=s_As_D+s_Bs_C+5[(s_A+s_D)b_1+(s_B+s_C)b_2] \eqno(s2) $$
and
$$ S_4=s_As_Bs_Cs_D+s_Bs_C(s_A+s_D)b_1+s_As_D(s_B+s_C)b_2+ $$
$$ 3(s_As_B+s_Bs_D+s_Ds_C+s_Cs_A)b_1b_2+s_As_Db^2_2+s_Bs_Cb^2_1+ $$
$$ 13b_1b_2[(s_A+s_D)b_2+(s_B+s_C)b_1]. \eqno(s4) $$
Using the above expressions, and ignoring terms of order
$(b')^{2}$ and $b^2$ which do not contribute anything to the
flow equations, we find the sublattice magnetizations per site as
$$ m'_A=\Biggl\{
   e^{6\sqrt{3}J'-2K'}\sinh\Bigl[[(42\sqrt{3}+24)J'-14K'](b'_1+b'_2)\Bigr]+ $$
$$  e^{-6\sqrt{3}J'+2K'}\sinh\Bigl[[(42\sqrt{3}-24)J'-14K'](b'_2-b'_1)\Bigr]
  \Biggr\}\times $$
$$ \Biggl\{
  e^{6\sqrt{3}J'-2K'}\cosh\Bigl[[(42\sqrt{3}+24)J'-14K'](b'_1+b'_2)\Bigr]+$$
$$ e^{-6\sqrt{3}J'+2K'}\sinh\Bigl[[(42\sqrt{3}-24)J'-14K'](b'_2-b'_1)\Bigr]
  \Biggr\}^{-1} \eqno(m2) $$
for the small cluster, and
$$ m_A=\Biggl\{
   e^{(24\sqrt{3}+8)J-7K}\sinh\Bigl[[(72\sqrt{3}+40)J-22K](b_1+b_2)\Bigr]+$$
$$ 2e^{-K}\sinh\Bigl[(72\sqrt{3}J-24K)b_2+(40J-2K)b_1\Bigr]+$$
$$ e^{(-24\sqrt{3}+8)J+9K}\sinh\Bigl[[(72\sqrt{3}-40)J-26K](b_2-b_1)\Bigr]
   \Biggr\}\times $$
$$ \Biggl\{
   e^{(24\sqrt{3}+8)J-7K}\cosh\Bigl[[(72\sqrt{3}+40)J-22K](b_1+b_2)\Bigr]+$$
$$ 2e^{-K}\cosh\Bigl[(72\sqrt{3}J-24K)b_2+(40J-2K)b_1\Bigr]+$$
$$ 2e^{-K}\cosh\Bigl[(72\sqrt{3}J-24K)b_1+(40J-2K)b_2\Bigr]+$$
$$ 2e^{-8J+K}+
   e^{(-24\sqrt{3}+8)J+9K}\cosh\Bigl[[(72\sqrt{3}-40)J-26K](b_2-b_1)\Bigr]
   \Biggr\}^{-1} \eqno(m4) $$
for the large cluster (where we have set $h'=h=0$.) Differentiation of
eqs. \(m2) and \(m4) with respect to $b'_1$, $b'_2$ and $b_1$, $b_2$ leads
to the flow equations \(mfl1) and \(mfl2) of the main text. The
zero-field susceptibilities needed for the calculation of the
magnetic exponent read as
$$ \chi'_{*}={{e^{6\sqrt{3}J_*-2K_*}}\over
				   {\cosh(6\sqrt{3}J_*-2K_*)}} \eqno(chipstar)
$$
and
$$ \chi_{*}={{4\Bigl[e^{(24\sqrt{3}+8)J_*-7K_*}+e^{-K_*}\Bigr]}\over
{e^{(24\sqrt{3}+8)J_*-7K_*}+4e^{-K_*}+2e^{-8J_*+K_*}+
 e^{(-24\sqrt{3}+8)J_*+9K_*}}}. \eqno(chistar) $$

\endpage
\references

\refis{mfrg} J.O. Indekeu, A. Maritan, and A.L. Stella, J. Phys. A {\bf{15}},
L291 (1982).

\refis{review} For a recent review see: A. Croes and J.O. Indekeu, Mod. Phys.
Lett. B {\bf{7}}, 699 (1993).

\refis{stella} J.O. Indekeu, A.L. Stella, and L. Zhang, J. Phys. A {\bf{17}},
L341 (1984).

\refis{slotte} P.A. Slotte, J. Phys. A {\bf{20}}, L177 (1987).

\refis{indekeu} J.O. Indekeu, A. Maritan, and A.L. Stella, Phys. Rev. B
{\bf{35}}, 305 (1987).

\refis{wagner} C.N. Likos, K.R. Mecke, and H. Wagner, J. Chem. Phys.
{\bf{102}},
9350 (1995).

\refis{oliv}  P.M.C. de Oliveira and F.C. S\'a Baretto, J. Stat. Phys.
{\bf{57}}, 53 (1989).

\refis{at} J. Ashkin and E. Teller, Phys. Rev. {\bf{64}}, 178 (1943).

\refis{plas} J.A. Plascak, Physica A {\bf{183}}, 563 (1992).

\refis{nauenberg} M. Nauenberg and B. Nienhuis, Phys. Rev. Lett. {\bf{33}},
944 (1974).

\refis{vanleeuwen} J.M.J. van Leeuwen, Phys. Rev. Lett. {\bf{34}}, 1056 (1975).

\refis{domb} C. Domb in {\it{Phase Transitions and Critical Phenomena}},
edited by C. Domb and M.S. Green (Academic, London, 1974), Vol. 3.

\refis{legui} J.C. Le Guillou and J. Zinn-Zustin, Phys. Rev. B {\bf{21}},
3976 (1980).

\refis{mcrg} G.S. Pawley, R.H. Swendsen, D.J. Wallace, and K.G. Wilson,
Phys. Rev. B {\bf{29}}, 4030 (1984).

\refis{hadwiger} H. Hadwiger, {\it{Vorlesungen \"uber Inhalt, Oberfl\"ache
und Isoperimetrie}} (Springer, Berlin, 1957).

\refis{weil} W. Weil in {\it{Convexity and its Applications}}, edited by
P.M. Gruber and J.M. Wills (Birkh\"auser, Basel, 1983)

\refis{remark} In \ref{wagner} above, the model was also exclusively
studied for the case $h_3>0$. The reason for that choice was that
$h_3 > 0$ puts large statistical weights on configurations with
negative Euler characteristic and thus encourages the stability of
phases with a random, bicontinuous nature such as microemulsions.

\refis{remark2} It has to be expected that {\it{only part}} of the
`critical line' can be preempted since, for small values of $h_3$ the
model is a slightly perturbed Ising model and the order of the phase
transition remains unchanged.

\refis{degennes} P.G. de Gennes and C. Taupin, J. Phys. Chem. {\bf{86}},
2294 (1982).

\refis{mecke} K.R. Mecke, Z. Phys. B {\bf{97}}, 379 (1995).

\refis{first} Th. Niemeijer and J.M.J. van Leeuwen in {\it{Phase Transitions
and Critical Phenomena}}, edited by C. Domb and M.S. Green (Academic,
London 1976), Vol. 6.

\endreferences
\endpage
TABLE I. The $T>0$-fixed points of the RG flows for the {\sq}-Ising
model with crossing bonds presented in the text [eqs. \(flis1) and \(flis2)]
along with the corresponding critical exponents.
\bigskip

\baselineskip=15pt  
\halign{
# \hfil &
# \hfil &
# \hfil &
# \hfil &
\hfil # \cr
\noalign {\hrule}
\noalign {\smallskip}
\noalign {\hrule}
\noalign {\smallskip}
& $\qquad (J_*,K_*)\qquad$
& $\qquad y_1\qquad$
& $\qquad y_2\qquad$
& $\qquad 2y_h-d$ \cr
\noalign {\smallskip}
\noalign {\hrule}
\noalign {\smallskip}
$C_1\qquad$ & $(0.1590, 0.1499)$ & $\;\;0.760$ & $\;-1.023$ & $0.974$ \cr
\noalign {\medskip}
$C_2\qquad$ & $(-0.1590,0.1499)$ & $\;\;0.760$ & $\;-1.023$ &$-0.037$ \cr
\noalign {\medskip}
$C_3\qquad$ & $(0,0.3465)$ & $\;\;1.073$ & $\;\;\;\;0.600$ & $0.830$ \cr
\noalign {\medskip}
$P\qquad$ & $(0,0)$ & $-1.170$ & $\;-0.830$ & $0$ \cr
\noalign {\smallskip}
\noalign {\hrule}
\noalign {\smallskip}
\noalign {\hrule}}
\endpage
TABLE II. The $T>0$-fixed points of the RG flows for the {\sc}-Ising
model with crossing bonds
and the corresponding exponents.
\bigskip

\baselineskip=15pt  
\halign{
# \hfil &
# \hfil &
# \hfil &
# \hfil &
\hfil # \cr
\noalign {\hrule}
\noalign {\smallskip}
\noalign {\hrule}
\noalign {\smallskip}
& $\qquad (J_*,K_*)\qquad$
& $\qquad y_1\qquad$
& $\qquad y_2\qquad$
& $\qquad 2y_h-d$ \cr
\noalign {\smallskip}
\noalign {\hrule}
\noalign {\smallskip}
$C_1\qquad$ & $(0.0629, 0.0580)$ & $\;\;0.506$ & $\;-0.424$ & $0.552$ \cr
\noalign {\medskip}
$C_2\qquad$ & $(-0.0629,0.0580)$ & $\;\;0.506$ & $\;-0.424$ &$-0.056$ \cr
\noalign {\medskip}
$C_3\qquad$ & $(0,0.0912)$ & $\;\;0.443$ & $\;\;\;\;0.344$ & $0.376$ \cr
\noalign {\medskip}
$P\qquad$ & $(0,0)$ & $-0.966$ & $\;-0.377$ & $0$ \cr
\noalign {\smallskip}
\noalign {\hrule}
\noalign {\smallskip}
\noalign {\hrule}}
\endpage
TABLE III(a). The $T>0$-fixed points of the RG flows for the {\bcc}-Ising
model with crossing bonds and the
corresponding exponents obtained from the tetrahedron $\to$ bond
mapping.
\bigskip

\baselineskip=15pt  
\halign{
# \hfil &
# \hfil &
# \hfil &
# \hfil &
\hfil # \cr
\noalign {\hrule}
\noalign {\smallskip}
\noalign {\hrule}
\noalign {\smallskip}
& $\qquad (J_*,K_*)\qquad$
& $\qquad y_1\qquad$
& $\qquad y_2\qquad$
& $\qquad 2y_h-d$ \cr
\noalign {\smallskip}
\noalign {\hrule}
\noalign {\smallskip}
$C_1\qquad$ & $(0.0781, 0.0791)$ & $\;\;0.653$ & $\;-0.848$ & $0.724$ \cr
\noalign {\medskip}
$C_2\qquad$ & $(-0.0781,0.0791)$ & $\;\;0.653$ & $\;-0.848$ &$-0.061$ \cr
\noalign {\medskip}
$C_3\qquad$ & $(0,0.2027)$ & $\;\;0.901$ & $\;\;\;\;0.650$ & $0.788$ \cr
\noalign {\medskip}
$P\qquad$ & $(0,0)$ & $-0.667$ & $\;-0.789$ & $0$ \cr
\noalign {\smallskip}
\noalign {\hrule}
\noalign {\smallskip}
\noalign {\hrule}}
\endpage
TABLE III(b). Same as in Table III(a), but using
the parallelogram $\to$ bond
mapping.
\bigskip

\baselineskip=15pt  
\halign{
# \hfil &
# \hfil &
# \hfil &
# \hfil &
\hfil # \cr
\noalign {\hrule}
\noalign {\smallskip}
\noalign {\hrule}
\noalign {\smallskip}
& $\qquad (J_*,K_*)\qquad$
& $\qquad y_1\qquad$
& $\qquad y_2\qquad$
& $\qquad 2y_h-d$ \cr
\noalign {\smallskip}
\noalign {\hrule}
\noalign {\smallskip}
$C_1\qquad$ & $(0.0994, 0.0494)$ & $\;\;0.426$ & $\;-0.800$ & $0.466$ \cr
\noalign {\medskip}
$C_2\qquad$ & $(-0.0994,0.0494)$ & $\;\;0.426$ & $\;-0.800$ &$-0.026$ \cr
\noalign {\medskip}
$C_3\qquad$ & $(0,0.2027)$ & $\;\;0.846$ & $\;\;\;\;0.650$ & $0.788$ \cr
\noalign {\medskip}
$P\qquad$ & $(0,0)$ & $-0.320$ & $\;-0.789$ & $0$ \cr
\noalign {\smallskip}
\noalign {\hrule}
\noalign {\smallskip}
\noalign {\hrule}}
\endpage
TABLE IV. The $T>0$-fixed points of the RG flows for the morphological
Hamiltonian and the corresponding exponents.
\bigskip

\baselineskip=15pt  
\halign{
# \hfil &
# \hfil &
# \hfil &
# \hfil &
\hfil # \cr
\noalign {\hrule}
\noalign {\smallskip}
\noalign {\hrule}
\noalign {\smallskip}
& $\qquad (J_*,K_*)\qquad$
& $\qquad y_1\qquad$
& $\qquad y_2\qquad$
& $\qquad 2y_h-d$ \cr
\noalign {\smallskip}
\noalign {\hrule}
\noalign {\smallskip}
$C_1\qquad$ & $(1.656\times 10^{-2}, 5.306\times 10^{-2})$
& $\;\;0.691$ & $\;-0.740$ & $0.646$ \cr
\noalign {\medskip}
$C_2\qquad$ & $(1.407\times 10^{-2}, 9.858\times 10^{-2})$
& $\;\;0.691$ & $\;-0.610$ &$-0.029$ \cr
\noalign {\medskip}
$C_3\qquad$ & $(2.785\times 10^{-2}, 1.447\times 10^{-1})$
& $\;\;0.728$ & $\;\;\;\;0.564$ & $0.516$ \cr
\noalign {\medskip}
$P\qquad$ & $(0,0)$ & $-0.667$ & $\;-0.789$ & $0$ \cr
\noalign {\smallskip}
\noalign {\hrule}
\noalign {\smallskip}
\noalign {\hrule}}
\endpage
\figurecaptions

FIG. 1. The $N'=2$ and $N=4$ clusters used for the remormalization of the
{\sq}-Ising Hamiltonian with crossing bonds. The fluctuating clusters are
shown by solid lines. The filled dots form the $A$-sublattice and the
open ones the $B$-sublattice.

FIG. 2. RG flows of the {\sq}-Ising Hamiltonian with crossing bonds.
The critical line is denoted by solid lines.

FIG. 3. A local arrangement of the {\bcc}-lattice and the
{\bcc} Wigner-Seitz unit cell.

FIG. 4. RG flows of the {\bcc} morphological Hamiltonian; (a) throughout
the subspace $h_3 \geq 0$ and (b) in more detail, in the neighborhood of
the critical points. The lines along which the critical points $C_1$ and
$C_2$ attract are denoted solid. The dashed line is the locus of
vanishing inverse susceptibility from the simple mean-field approximation.
The mean-field tricritical point is indicated by the dot.
The filled triangles denote values of the
energy parameters for which the mean-field approximation yields
three-phase coexistence between oil-rich, water-rich and a middle,
random, bicontinuous microemulsion.

\endfigurecaptions
\end